\documentclass[aps,preprint,onecolumn,eqsecnum]{revtex4}
\usepackage{dcolumn}
\usepackage{bm}
\usepackage{amssymb}
\usepackage{amsmath}
\usepackage{amsfonts}

\usepackage[]{graphicx}
\begin{document}
\bibliographystyle{prsty}
\title{How to justify Born's rule using the pilot wave theory of de Broglie?}
\author{ Aur\'elien Drezet $^{1}$}
\address{(1) Univ. Grenoble Alpes, CNRS, Institut N\'{e}el, F-38000 Grenoble, France
}
\begin{abstract}
 In this article we discuss few new derivations of the so called Born's rule for quantum probability in the context of the pilot wave theory proposed by de Broglie in 1927. 
[This is a  corrected version of my article published in Annales de la Fondation Louis de Broglie in 2017. ] 
\end{abstract}

 \maketitle
\section{Introduction}
\indent Louis de Broglie anticipated already in the 1920's that the usual Copenhagen interpretation will not be the final view  concerning quantum mechanics.  From the start he proposed, following great classical masters like Albert Einstein and Gustav Mie, to attribute a dual nature to any quantum system.  In his double solution program~\cite{deBroglie,livre}, he postulated that a quantum object should be seen as a localized singularity of a field  producing a guiding wave piloting its motion.  The singularity was replaced progressively by the concept of soliton as a localized stable wavelet  (i.e. solution of a non linear equation) propagating as a whole and reproducing the rules of quantum mechanics.  The program was so ambitious that de Broglie could only find some particular  examples (like a free particle in uniform motion)  to illustrate the concept of double solution. Therefore, in 1927 during the 5$^{th}$ Solvay~\cite{Solvay} he preferred to restrict the discussion to the so called pilot wave dynamics which was later rediscovered by David Bohm in 1952~\cite{Bohm1952}. The approach was strongly criticized by many, including Heisenberg and Pauli, who coined the pilot wave  `metaphysical' or `surrealistic' (this is discussed in \cite{Drezetbook}). The name `hidden variable' is generally associated with the approach of de Broglie and Bohm. Actually, it is interesting to see that in many other fields  of research   hidden variables are more or less accepted or tolerated. Think for example about Quarks  which can not be seen directly and can not be isolated. Think about the concept of black-hole in general relativity: Nobody or nothing could communicate about the internal region of the black-hole beyond the horizon. Clearly, these examples are hidden worlds. However, these examples are accepted by the community because they can be useful  even if all the consequences are not directly testable in the Poperian sense.  Therefore, in my opinion rejecting Louis de Broglie double solution or pilot wave approaches because these include hidden variables is not fair and not very pertinent.  What is important is to see if the double solution program of de Broglie can lead to observable consequences. What is moreover known is that the pilot wave dynamics is a clear ontological model for reproducing all the predictions of the usual quantum mechanics (at least in the non relativistic regime). This is already a great achievement. However, the pilot wave is not unique and we could imagine alternative guiding laws for reproducing quantum mechanics. Therefore, a foundation is missing and de Broglie worked hard to find such a foundation without a clear success.\\
\indent The question is thus can we do better?  De Broglie and Bohm  with different strategies thought that two domains of physics could be interesting for future investigations. The first one, considers the high energy regime where particles can be created and  annihilated.  Despite many attempts since the 1970's no  progress has been obtained for going beyond the standard model which was inaugurated by  the paradigmatic QED  after  the second world-war. The so called string theory approach did not bring  anything physical until now  and may be  it is really time to consider the double solution program as a real alternative to go beyond the standard model. A theory of solitons could potentially lead to a mathematical justification for the mass spectrum of particles and this goes clearly beyond the current understanding of the standard  model and could really motivate a new paradigm.  The second big problem, is the justification of statistical predictions in quantum mechanics. What is really  interesting  with the pilot wave model  is that, as a deterministic theory, i.e.  like  Newtonian dynamics, it contains, in principle,  enough ingredients to justify the presence of probability in the quantum world.  Probability is indeed a  mysterious concept since it relies on a precise quantification of the elusive notions of chance and randomness. Classically, with Maxwell or Laplace this is associated with ignorance and a lack of information. Therefore, probability are not fundamental and rely on some contingency or postulate on the initial conditions. This is also the case in the pilot-wave framework.  More precisely, in classical statistical physics the Liouville theorem allows us to give a dynamical foundation  to thermodynamics  if we include some postulates about randomness and molecular chaos like it was done by  Maxwell, Boltzmann, and Gibbs.  The pilot wave interpretation inherits these advantages and defects of classical Newtonian dynamics. Indeed, many discussions about which postulates are necessary to be added to the pure dynamical law for justifying the statistics observed in the physical world continue since the 19$^{th}$ century. These problems and discussions are  clearly surviving in the context of the pilot wave approach (for a recent review of different approaches on this domain see ref.~\cite{Callender}) and therefore the question of how to justify quantum rules for probability is in large  part still open (see the analysis  by Pauli~\cite{Pauli} and Keller~\cite{Keller}). In the present work we will discuss some alternative justifications of the so  called Born's rule for quantum  probability. These deductions could shed new light on the conditions of validity of the so called quantum equilibrium regime.  Furthermore, following  A. Valentini \cite{Valentini} we believe that the studies of non-equilibrium states, probably at the beginning of the Universe, could bring some important insight on the foundation of quantum mechanics. Therefore, the present work is only a contribution to this important problem which will, we think, continue to motivate many researchers.                                      
\section{General properties of the de Broglie-Bohm dynamics}
In order to understand how probabilities enter into the de
Broglie-Bohm framework it is crucial to clarify the role of dynamics
in this theory. The de Broglie-Bohm mechanics starts with the
Schr\"{o}dinger equation and for the present work we will limit the analysis to the non relativistic version involving $N$
interacting spinless point-like particles labeled by $i$, with
$i=1,...,N$. For the same reason, but still without loosing
generality, we will also omit from the theory the potential vector
$\mathbf{A}(\mathbf{x},t)$. Now, Schr\"{o}dinger's equation in the
spatial-coordinate representation reads
\begin{eqnarray}
i\frac{\partial \Psi(\mathbf{x}_1,...,\mathbf{x}_N,t)}{\partial
t}=-\sum_{i=1}^{i=N}\frac{\boldsymbol{\nabla}_i^2}{2
m_i}\Psi(\mathbf{x}_1,...,\mathbf{x}_N,t)\nonumber\\+
V(\mathbf{x}_1,...,\mathbf{x}_N,t)\Psi(\mathbf{x}_1,...,\mathbf{x}_N,t)
\end{eqnarray}
where $V(\mathbf{x}_1,...,\mathbf{x}_N,t)$ is the interaction
potential arising either form external or internal coupling, $m_i$
denote the particle masses and
$\boldsymbol{\nabla}_i=\frac{\partial}{\partial
\mathbf{x}_i}$ are the gradient operators. We will use very often the super-vector notation $\textbf{X}(t)=[\mathbf{x}_1(t),...,\mathbf{x}_N(t)]$.\\
\indent Using the polar representation
$\Psi(\mathbf{X},t)=a(\mathbf{X},t)
e^{iS(\mathbf{X},t)}$ with $a$ and $S$ real
valued numbers, we get the two de Broglie-Madelung equations:
\begin{eqnarray}
-\frac{\partial S(\mathbf{X},t)}{\partial
t}=-\sum_{i=1}^{i=N}\frac{(\boldsymbol{\nabla}_i
S(\mathbf{X},t))^2}{2 m_i}+
V(\mathbf{X},t)-\sum_{i=1}^{i=N}\frac{1}{2m_i}\frac{\boldsymbol{\nabla}_i^2
a(\mathbf{X},t)}{a(\mathbf{X},t)},\label{madelung1}
\end{eqnarray}and
\begin{eqnarray}
-\frac{\partial a^2(\mathbf{X},t)}{\partial
t}=-\sum_{i=1}^{i=N}\boldsymbol{\nabla}_i[a^2(\mathbf{X},t)
\cdot\frac{\boldsymbol{\nabla}_i
S(\mathbf{X},t)}{m_i}].\label{madelung2}
\end{eqnarray}
Starting with Eq.~\ref{madelung1}, the analogy with  the classical
Hamilton-Jacobi equation for an ensemble of  $N$ point-like objects
interacting through  $V$ is immediate.  This motivates the definition
of particle trajectory  for the quantum system by introducing the de
Broglie-Bohm velocity for the $N$ points as:
\begin{eqnarray}
\mathbf{v_i(t)}=\frac{d\mathbf{x}_i(t)}{dt}=\frac{\boldsymbol{\nabla}_i
S(\mathbf{X}(t),t)}{m_i}]
=\frac{1}{m_i}\textrm{Im}[\frac{\boldsymbol{\nabla}_i\Psi(\mathbf{X}(t),t)}{\Psi(\mathbf{X}(t),t)}].\label{velocity}
\end{eqnarray}
These are first order equations which can be integrated directly  to
give the particle trajectories in the de Broglie-Bohm mechanics as
\begin{eqnarray}
dt=m_i\frac{dx_i}{\partial_{x_i}S}=m_i\frac{dy_i}{\partial_{y_i}S}=m_i\frac{dz_i}{\partial_{z_i}S}.
\end{eqnarray}
Now, as it was done by both de Broglie and Bohm, this dynamics can
also be written in a second-order Newtonian form:
\begin{eqnarray}
\frac{d\mathbf{v_i(t)}}{dt}=-\boldsymbol{\nabla}_i[V(\mathbf{X}(t),t)+Q(\mathbf{X}(t),t)]\label{Newton}
\end{eqnarray} where $Q$ is the quantum potential defined as
\begin{eqnarray}
Q(\mathbf{X},t)=-\sum_{i=1}^{i=N}\frac{1}{2m_i}\frac{\boldsymbol{\nabla}_i^2
a(\mathbf{X},t)}{a(\mathbf{X},t)}.\label{qpotential}
\end{eqnarray} This shows that the de Broglie-Bohm mechanics is highly non
classical since the quantum potential is in general acting in a
contextual and nonlocal way on the particles. We also emphasize that
$S(\mathbf{X},t)$ is not in general a univocal
function of space since a phase is defined modulo $2\pi$. When the
wave function cancels  (i.e. at nodes) the gradient of $S$ becomes
undefined and this could imply the existence of line singularities,
i.e., vortices. A closed integral  along a loop $(C)$ surrounding
such a singularity defines a Bohr-Sommerfeld like quantization
condition $\oint_{(C)} \sum_i\mathbf{v}_i\cdot d\mathbf{x}_i=2\pi
n_i$ through the integers $n_i=0,\pm 1,\pm 2,...$. This implies that
we should be cautious when trying to integrate directly
Eq.~\ref{Newton} because the Newtonian equations have too many
solutions not all satisfying this quantization rule and therefore
not always akin to Shr\"{o}dinger's equation. Actually, since
$S(\mathbf{X},t)$ is given by Schr\"{o}dinger's
equation the dynamic is entirely defined by the first-order
Eq.~\ref{velocity} also called the
guidance equation.\\
\indent The present work will be more interested into the second
Madelung Eq.~\ref{madelung2} associated with a conservation
condition. In usual quantum mechanics Eq.~\ref{madelung2} is
reminiscent of the probability conservation law
$-\partial_t\rho=\sum_i\boldsymbol{\nabla}_i\cdot\mathbf{J}_i$ where
$\rho=|\Psi|^2=a^2$ is the space density of probability and
$\mathbf{J}_i=\rho\mathbf{v}_i$ is the probability current. Here, in
the context of the de Broglie-Bohm mechanics the same should be true
but now probability will have a different interpretation because the
theory is deterministic contrarily to the orthodox view.  Before to
come to that problem, which constitutes the core of the present
article, we emphasize that Eq.~\ref{madelung2} actually must have
also a dynamical meaning in the theory  independently of any
probabilistic considerations. This point which is very similar to
Liouville's theorem in classical dynamics can be understood by
introducing the Lagrange derivative
$d/dt=\partial_t+\sum_i\mathbf{v}_i\cdot\boldsymbol{\nabla}_i$ which
applied on $a^2$ gives
\begin{eqnarray}
\frac{d}{dt}\ln{(a^2(\mathbf{X}(t),t))}=-\sum_i\boldsymbol{\nabla}_i\cdot\mathbf{v}_i(\mathbf{X}(t),t).\label{flow1}
\end{eqnarray}
This equation can be  formally integrated and leads to
\begin{eqnarray}
a^2(\mathbf{X}(t_{out}),t_{out})=a^2(\mathbf{X}(t_{in}),t_{in})
\cdot e^{-\sum_i\int_{t_{in}}^{t_{out}}
dt'\boldsymbol{\nabla'}_i\cdot\mathbf{v}_i(t')}\label{liouville}
\end{eqnarray} where $\mathbf{X}(t_{in})$ and $\mathbf{X}(t_{out})$
are the spatial coordinates respectively associated with the initial
and final points along a trajectory and where the integration is
made along this trajectory between the two times $t_{in}$ and
$t_{out}$. We emphasize that the polar form $\Psi=ae^{iS}$ also
means that we can write
\begin{eqnarray}
\Psi(\mathbf{X}(t_{out}),t_{out})
=\Psi(\mathbf{X}(t_{in}),t_{in})
e^{-\sum_i\int_{t_{in}}^{t_{out}}
dt'\boldsymbol{\nabla'}_i\cdot\mathbf{v}_i(t')/2}\nonumber\\
\cdot e^{i\int_{t_{in}}^{t_{out}}
dt'L(\mathbf{X}(t'),d\mathbf{X}(t')/dt',t')}
\end{eqnarray} where $L$ is the Lagrange function
\begin{eqnarray}
L(\mathbf{X}(t),\frac{\mathbf{X}(t)}{dt})
=\sum_i\frac{m_i\mathbf{v}_i^2}{2}-V(\mathbf{X},t)-Q(\mathbf{X},t).
\end{eqnarray}
What is however important here is that the de Broglie-Bohm dynamical
laws imply that a comoving elementary volume $\delta^{3N}
\textbf{X}(t)=\int_{\delta^{3N} \textbf{X}(t)}\prod_{i=1}^{i=N}d^3\mathbf{x}_i(t)$
centered on the point $\textbf{X}(t)=[\mathbf{x}_1,...,\mathbf{x}_N]$ in the
configuration space should satisfy the condition:
\begin{eqnarray}
\frac{d}{dt}\ln{(\delta^{3N}
\textbf{X}(t))}=+\sum_i\boldsymbol{\nabla}_i\cdot\mathbf{v}_i(\mathbf{X}(t),t).\label{flow2}
\end{eqnarray} This can be obtained directly by observing that from the conservation law  Eq.~\ref{madelung2} the product
$a^2(\textbf{X}(t),t)\delta^{3N} \textbf{X}(t)$ should be a conserved quantity during
the motion, i.e., $\frac{d}{dt}(a^2(\textbf{X}(t),t)\delta^{3N} \textbf{X}(t))=0$ which
together with Eq.~\ref{flow1} implies Eq.~\ref{flow2}. Alternatively
we can obtain Eq.~\ref{flow2} by considering how the points on the
boundary of a small volume evolves in times in analogy with
hydrodynamics. Since Eq.~\ref{flow2} can also be formally integrated
as\begin{eqnarray} \delta^{3N} \textbf{X}(t_{out})=  \delta^{3N} \textbf{X}(t_{in})
\cdot e^{+\sum_i\int_{t_{in}}^{t_{out}}
dt'\boldsymbol{\nabla'}_i\cdot\mathbf{v}_i(t')}\label{liouville2}
\end{eqnarray} this clearly implies that the de Broglie-Bohm fluid is not
incompressible and that the density of points in $\delta^{3N} \textbf{X}(t)$
is changing with time along trajectories.
\section{Probability and Born's rule in the pilot wave formalism: Geometry and symmetry }
\indent From now it is clear that the quantity $\delta \Gamma(t)=|\Psi|^2(\textbf{X}(t),t)\delta^{3N}
\textbf{X}(t))$  plays the role of a conserved measure in the dynamics since we have along any path  
$\frac{d}{dt}\delta \Gamma(t)=0$. As it was stated by Poincar\'{e}, in the context of  Newtonian physics, such a conserved invariant function is naturally expected to play a crucial role in any probabilistic interpretation of the pilot wave. Actually, if we want to have the most general probabilistic law following the analogy with Liouville's theorem one must introduce the density of probability $\rho(\mathbf{X}(t),t)=f(\mathbf{X}(t),t)|\Psi|^2(\textbf{X}(t),t)$   such as the conservation law
\begin{eqnarray}
-\frac{\partial\rho(\mathbf{X}(t),t)}{\partial
t}=-\sum_{i=1}^{i=N}\boldsymbol{\nabla}_i[\rho(\mathbf{X}(t),t))\mathbf{v}_i(\mathbf{X}(t),t)].\label{madelung2bis}
\end{eqnarray} occurs. This together with the law $\frac{d}{dt}\delta \Gamma(t)=0$ leads to the fundamental requirement: 
 \begin{eqnarray}
\frac{d}{dt}f(\mathbf{X}(t),t)=0 \label{liouvilian}
\end{eqnarray} where $f(\mathbf{X}(t),t)$ can be thus interpreted, in complete analogy with Liouville's theorem, as a probability density with respect to the measure $\Gamma$. 
Furthermore, we have the conservation law (after normalization) 
\begin{eqnarray}
\int \rho(\mathbf{X}(t),t)d^{3N}
\textbf{X}(t))=\int_\Gamma f(\mathbf{X}(t),t)d\Gamma(t)=\int_\Gamma d\Gamma(t)=\Gamma.\label{norm}
\end{eqnarray}
\indent The fundamental issue is thus to determine the distribution $f(\mathbf{X}(t),t)$. De Broglie originally  postulated in 1926-1957 that the  microcanonical  choice $f(\mathbf{X}(t),t)=1$ was the most natural but the proof is far from being obvious . Clearly,  the condition $f(\mathbf{X}(t),t)=1$ leads to the usual Born's law $\rho(\mathbf{X}(t),t)=|\Psi|^2(\textbf{X}(t),t)$ but a clear theoretical justification of the choice was lacking at that time. However, de Broglie motivated his choice by a physical example where free plane waves overlap and thus interact.  Each of the plane waves have a constant amplitude so that equiprobability was the most natural hypothesis.  Since $|\Psi|^2(\textbf{X}(t),t)=Const.$ for such waves   this leads naturally to $f$ constant as well and therefore in agreement with the definition Eq.~\ref{norm} to $f(\mathbf{X}(t),t)=1$. Moreover, the Liouville theorem ensures that the condition $f(t)=1$ will be valid for all time. Therefore, knowing that this occurs before the interaction of the plane waves will guarantee that it will still be true after the interaction. Therefore, the physical deduction of de Broglie, even if not very general, was very intuitive.  In later works~\cite{livre} de Broglie attempted to motivate a justification of Born's rule using  the ergodic or pseudo-ergodic principle which played an important historical role in early Boltzmann's studies. However, ergodicity means that we can replace ensemble average by time average and it is not clear which time scale should be used here.  The typical recurrence  (Poincar\'e) time in a large system is so huge (as Boltzmann showed in reply to Zermelo) that it is in general meaningless. Also, ergodicity is associated with classical equilibrium which is time independent. This is however not the case in general in quantum physics where Born's rule is also used for time dependent problem (the meaning of an hypothetical ergodic relation $dt/T=d\Gamma(\textbf{X},t)/\Gamma$, with $dt$ the time spent  by the system in the cell $d\Gamma(\textbf{X},t)$ and $T$ the total Poincar\'e time, is in general not clear since $d\Gamma(\textbf{X},t)$ depends explicitly of $t$). If ergodicity is used it could be only on some very specific conditions (may be at beginning of the Universe). More generally, ergodicity should be considered with caution and in fact Boltzmann used it principally as a guideline for motivating its `typicality' reasoning  with Bernoulli sequences.       \\
\indent In the 1990's D.~D\"{u}rr, S.~Goldstein, N.~Zangh\`{\i}~\cite{Goldstein} used such a reasoning based on the Boltzmann concept of typicality~\cite{Boltzmann,Durr,Lebowitz} to justify Born's rule. The reasoning relies on the Bernoulli (weak) law of large numbers ensuring that the phase volume occupied by `non typical' points in a large Gibbs space $\Gamma_M=\Gamma^M$ tends to vanish in the limit where the number of identical systems $M$ in this Gibbs ensemble goes to infinity. While the complete reasoning is too long to be discussed here in details  we should sketch a derivation based on the reasoning proposed by Boltzmann in 1877~\cite{Boltzmann}.   We start with a factorizable wave function $\Psi_M(\textbf{X}_1,...,\textbf{X}_N,t)=\bigotimes_{k=1}^{k=M}\psi(\textbf{X}_k, t)$ where $M$ is a large number and where the  `small' systems  labeled by $k$ are all characterized by the same wave function of the argument $\textbf{X}=[\mathbf{x}_1,...,\mathbf{x}_N]$ which defines a phase space $\Gamma$ for the $3N$ particle coordinates. The large system constitutes at the limit $M\rightarrow+\infty$ a kind of Gibbs ensemble and is characterized by a phase space $\Gamma_M=\Gamma^M$ as defined before. Now, we expand $\psi$ as a sum of non overlapping wave function $\psi_\alpha$ defined in the disjoint cells $\alpha$ such as $\Gamma=\sum_\alpha \Gamma_\alpha$ and where $\Gamma_\alpha=\int_{\alpha}|\psi(\textbf{X},t)|^2d^{3N}\textbf{X}=|\psi_\alpha|\delta^{3N}\textbf{X}_\alpha$ integrated on the cells defines a  `coarse graining' distribution.  We now use the multinomial expansion of $\Gamma_M=(\sum_\alpha \Gamma_\alpha)^M$    
and write \begin{eqnarray}
\Gamma_M =\sum_{\{m_{\alpha}\}}\frac{M!}{\prod_{\alpha}m_{\alpha}!}\prod_{\alpha}(\Gamma_\alpha)^{m_{\alpha}}=\sum_{\{m_{\alpha}\}}\Gamma_M(\{m_{\alpha}\})\label{complexion}
\end{eqnarray} 
where as usual $m_\alpha$ is the number of sub-systems present in the cell $\alpha$ such as $M=\sum m_\alpha$ and the sum in Eq.~\ref{complexion} is on the complete set of complexions.
In the $M\rightarrow+\infty$ limit the complexion coefficients of this
multinomial probability law $W(\{m_\alpha\})=M!/\prod_{\alpha}m_{\alpha}!$ are in general huge numbers for almost all configurations $\{m_\alpha\}$. Furthermore, this Functional is very well peaked around a
particular subset $\{\tilde{m}_{\alpha}\}$. To determine  this subset it is convenient after Boltzmann~\cite{Boltzmann,Durr}  to use the entropy $S_B(\Gamma_M(\{m_{\alpha}\}))=\ln{\Gamma_M(\{m_\alpha\})}$. Therefore, the distribution $\{\tilde{m}_{\alpha}\}$ corresponds to an optimum and we have near this point the variation $\delta S_B(\{\tilde{m}_{\alpha}\})=0$ with the
constraint $\sum_{\alpha}\tilde{m}_{\alpha}=M$. Using Stirling formula and the method of Lagrange multipliers we get the standard result  
\begin{equation}
\frac{\tilde{m}_{\alpha}}{M}=\frac{\Gamma_\alpha}{\Gamma}=\Gamma_\alpha\label{borns}
\end{equation}which is valid in the $M\rightarrow+\infty$ limit and constitutes Born's law (in agreement with conventions we use the normalization $\Gamma=1$). In other words if we define $\frac{\tilde{m}_{\alpha}}{M}=f_\alpha\Gamma_\alpha$  then Eq.~\ref{borns} indeed reads $f_\alpha=1$ in agreement with our definition of Born's rule. Importantly, using such a deduction we have also near the optimum  $\{\tilde{m}_{\alpha}\}$
\begin{eqnarray}
\frac{\Gamma_M(\{m_{\alpha}\})}{\Gamma_M}\simeq \frac{\sqrt{(2\pi
M)}}{\prod_\alpha\sqrt{(2\pi
\tilde{m}_\alpha)}}\exp{[-\frac{1}{2}\sum_{\alpha}\frac{\delta
m_{\alpha}^{2}}{\tilde{m}_\alpha}]},\label{gauss}
\end{eqnarray}  (where $\delta
m_{\alpha}=
m_{\alpha}-\tilde{m}_\alpha$) which  represents the Laplace-Gauss limit of the multinomial statistics defined before.  It is also clear from  Eq.~\ref{gauss} that the distribution will have
a typical width $\Delta m_{\alpha}\sim \sqrt{\tilde{m}_{\alpha}}$. We deduce
$\Delta m_{\alpha}/\tilde{m}_{\alpha}\sim 1/\sqrt{\tilde{m}_{\alpha}}\propto
1/\sqrt{M}$ which approaches zero asymptotically if $M$ tends to
infinity. This means that for the overwhelming majority of
possibilities Born's law should hold in agreement with Bernoulli's
(weak) law of large numbers.\\ \indent  This statement can be rigorously formulated using the Bienayme-Chebyshev inequality which reads here   
\begin{eqnarray}
\frac{\Delta\Gamma_M[|\frac{m_{\alpha}}{M}-\Gamma_\alpha|\geq \varepsilon\sqrt{\Gamma_\alpha(1-\Gamma_\alpha)}]}{\Gamma_M}\leq \frac{1}{\varepsilon^2 M}
\end{eqnarray} where $\varepsilon$ is any positive number. $\Delta\Gamma_M[...]=F(\alpha,M)$ is defined for a given $\alpha$ and contains all points satisfying the constraint $|\frac{m_{\alpha}}{M}-\Gamma_\alpha|\geq \varepsilon\sqrt{\Gamma_\alpha(1-\Gamma_\alpha)}$. Now, we can define a multivariate inequality by using  the relation $F(\alpha, \textrm{ and } {}\beta,M)\leq F(\alpha,M)+F(\beta,M)$. Therefore, we get 
\begin{eqnarray}
\frac{\Delta\Gamma_M[\{\frac{m_{\alpha}}{M}-\Gamma_\alpha|\geq \varepsilon\sqrt{\Gamma_\alpha(1-\Gamma_\alpha)}\}_{\Delta M}]}{\Gamma_M}\leq \frac{\Delta M}{\varepsilon^2 M}
\end{eqnarray}  where $\Delta M$ is the number of cells considered in the subset. Using the continuous limit $\delta^{3N}\textbf{X}_\alpha\rightarrow0$ we can define $\Delta M=\Delta^{3N}\textbf{X}/\delta^{3N}\textbf{X}_\alpha$ where the elementary volume is supposed to be the same for every $\alpha$ and where $\Delta^{3N}\textbf{X}$ is the total volume of the subset considered. Therefore, we finally get:
\begin{eqnarray}
\frac{\Delta\Gamma_M[|f(\textbf{X},t)-1|\geq \frac{\varepsilon}{\sqrt{\Delta^{3N}\textbf{X}|\psi(\textbf{X},t)|^2}}, \forall\textbf{X}\in\Delta^{3N}\textbf{X}]}{\Gamma_M}\leq \frac{1}{\varepsilon^2 M}
\label{weaklaw}\end{eqnarray} which constitutes a geometrical statement of the weak law of large numbers.  Clearly, whatever $\varepsilon>0$ it is always possible to find a $M$ sufficiently large such as $\frac{\Delta\Gamma_M}{\Gamma_M}\rightarrow 0$ asymptotically.\\
\indent The Boltzmann-Bernoulli strategy is remarkable but it contains \textit{in fine} necessarily some circularity since it relies on a concept of typicality or if we want of equiprobability in $\Gamma_M$ which, somehow, means that we accepted already  a condition like $f_M=1$ from the start in the large ensemble $\Gamma_M$. Of course, typicality (or may be plausibility) is intuitive since the choice $f_M=1$ corresponds to the simplest stationary ensemble. However, we point out that we could actually extend a bit the condition of validity of the typicality proof by considering that instead of $f_M=1$ a relatively smooth distribution such as $\Delta P_M=\int_{\Delta\Gamma_M} f_M d\Gamma_M$ represents the probability for the large system to be in the volume $\Delta\Gamma_M$ given in Eq.~\ref{weaklaw}. If we call $f_{M,\textrm{max}}$ the maximum value taken by $f_M$ in $\Delta\Gamma_M$ we have 
\begin{eqnarray}
\Delta P_M\leq f_{M,\textrm{max}} \Delta\Gamma_M\leq f_{M,\textrm{max}} \frac{\Gamma_M}{\varepsilon^2 M}\end{eqnarray}
which implies that $\Delta P_M$ tends to zero asymptotically for large $M$ if $f_{M,\textrm{max}}\Gamma_M $ is finite, in agreement with the weak law of large number .\\ 
Nevertheless, we now point out that the authors of the typicality methods generally state that their choice $f_M=1$ should not be confused with a probability measure since within  the typicality interpretation we do not require a Gibbs ensemble of Universes: only one is apparently needed. Still many authors are not convinced by such an explanation (this was already true at the time of Boltzmann, Zermelo, Loschmidt and Poincar\'e). This distribution $f_M$ could perhaps be interpreted in a Bayesian sense, i.e., as a subjective probability but we will not follow this strategy here (see Refs.~\cite{Everett1957,Drezet2} for a discussion in the context of the Many World Interpretation). 
Moreover, nothing forces us to believe in the typicality interpretation of a pure geometrical or combinatorial statement. Why a small (but not rigorously vanishing) volume in the $\Gamma_M$ space should have physically less importance than a larger one? Why our Universe could not be in a `maverick', i.e. atypical state? In such an extraordinary state $f_M$ could be extremely peaked in a region of atypicality  (i.e. such as $\Delta P_M$ does not cancel even if $M\rightarrow +\infty$). Furthermore,  we can argue that in the typicality framework the authors confuse the conclusion and the premise. Therefore, it seems that typicality is merely a tautological statement telling  that you should call `typical' what is already known to be `actual'.  Typicality would be in other words nothing else that an elegant mathematical characterization of what we know already to be the good result, i.e., $f_M=1$ or $f_M\sim 1$ because we experience it. Interestingly, Boltzmann wrote his famous article on the statistical entropy $S_B=\ln{\Gamma_M}$ in 1877~\cite{Boltzmann}, but only in his later book on gas dynamics~\cite{Gas1} in 1895-8 did he presented a unified view concerning his choice for $S_B$  and his earlier 1872 famous `H-theorem' based on kinetic equations~\cite{Gas2}. Both approaches lead to the same conclusion and clearly Boltzmann wanted to emphasize the fact that many complementary perspectives are probably needed to reach a high degree of confidence in the field of thermodynamics.\\
\indent In the present work we want to motivate further the justification $f=1$, or if we want $f_M=1$ without necessarily relying on typicality. Therefore we want to motivate a kinetic derivation of quantum equilibrium $f=1$ like Boltzmann did in 1872. This `proof' will be far from the ideal as any H-theorem proof is by the way. Still, we think that it could motivate further analysis.  We should first observe that several other methods based on different axiomatic can be used in thermostatistics to derive the equilibrium laws.  For instance, if we consider an isolated system with phase space $\Gamma$ we can define with Boltzmann and Gibbs the entropy 
\begin{eqnarray}
S_G=-\int_{\Gamma}
f\ln{f}d\Gamma
\end{eqnarray}  Following  the strategy developed by  Gibbs, Shannon and
Jaynes (but with different interpretations of  what should be a
probability) we could then try to find an optimum to the variational
problem $\delta S_G=-\int_{\Gamma} \delta
f(\ln{f}+1)d\Gamma=0$ when submitted to the constraint
$\int_{\Gamma} \delta fd\Gamma=0$. The Lagrange multiplier
method would give us directly the solution
$f_N=Const.=1$, i.e., the microcanonical ensemble
from which  Born's rule can be deduced.  This `proof'  of course relies on the concept of entropy and one could ultimately find again some circularity on the deduction. In particular since   $dS_G(t)/dt=0$ for all time we deduce that if $S_G$ reach the optimum $S_B=\ln{\Gamma}=\ln{(1)}=0$ at one time it will conserves this value at all time. There is therefore with such an interpretation a kind of miracle which fixes the quantum equilibrium and which can not be explained further (the reasoning is therefore very similar to the typicality approach).   Actually, it is probably impossible to give a perfect proof of the microcanonical law $f=1$ along such lines of thought and the reasoning in statistical physics are often guided by some symmetry consideration on equiprobability, i.e.,  strongly motivated by physical intuitions.\\
\indent It is important to observe that the stationary  distribution $f=1$ is actually stable  with respect to first-order fluctuations.  To see that we write $i\partial_t f=\mathcal{L}_\psi[f]$ the Liouville equation with
\begin{eqnarray}
\mathcal{L}_\psi[f]=-i\sum_i\mathbf{v}_i(\mathbf{X},t)\cdot\boldsymbol{\nabla}_i f
\end{eqnarray}
Now, if the wave function is disturbed by let say an external force we have the new wave function $\Psi=\Psi^{(0)}+\Psi^{(1)}$ and the velocity fields is changed accordingly such as $\mathcal{L}=\mathcal{L}^{(0)}+\mathcal{L}^{(1)}$. In this linear response theory\cite{Zwanzig} we get, after writing $f=f^{(0)}+f^{(1)}$, the two equations: $i\partial_t f^{(0)}=\mathcal{L}^{(0)}[f^{(0)}]$ and
 \begin{eqnarray} i\partial_t f^{(1)}=\mathcal{L}^{(0)}[f^{(1)}]+\mathcal{L}^{(1)}[f^{(0)}].\label{ders}\end{eqnarray}
 Moreover, if we use the stationary zero-order solution $f^{(0)}=1$  then Eq.~\ref{ders} reduces to  $i\partial_t f^{(1)}=\mathcal{L}^{(0)}[f^{(1)}]$. If we impose $f^{(1)}=0$ at a given time $t$ we deduce by iteration that $\partial_t f^{(1)}=0$, $\partial^2_t f^{(1)}=0$, etc.. at such time $t$. Using a Taylor expansion this means that $f^{(1)}=0$ is solution for all time and thus that the condition $f=1$ is not modified by the linear response theory. An equivalent deduction can be obtained using the   formal  Dyson expansion $f^{(1)}(t)=\mathcal{T}[e^{-i\int_{t_0}^t dt'\mathcal{L}_{t'}^{(0)}}]f^{(1)}(t_0)$ with $\mathcal{T}$ a time ordered product operator. This `derivation' again stresses the strong naturalness of the micro-canonical distribution $f=1$ in the pilot wave framework.   \\
\indent  We would like to suggest an other proof based on a version of the well known Gleason theorem  in quantum mechanics. The `poor man' version of Gleason's theorem that we use  here can be stated
like that (this version goes back to Everett~\cite{Everett1957}, but  the calculation details are ours): We
start from the postulated probability additivity
\begin{eqnarray}
g(\sum_\alpha |c_{\alpha}|^2)=\sum_\alpha g(|c_{\alpha}|^2)
\end{eqnarray} where the sum is over a complete set of wave function constituting a basis in the Hilbert space.
Choosing one particular $\alpha_0$ and taking the partial derivative
$\partial[...]/\partial(|c_{\alpha_0}|^2)$ on both sides we get on
the left hand side :
\begin{eqnarray}
\frac{\partial g(\sum_\alpha
|c_{\alpha}|^2)}{\partial(|c_{\alpha_0}|^2)}= \frac{dg(\sum_\alpha
|c_{\alpha}|^2)}{d(\sum_\alpha|c_{\alpha}|^2)}\frac{\partial(\sum_\alpha|c_{\alpha}|^2)}{\partial(|c_{\alpha_0}|^2)}=\frac{dg(\sum_\alpha
|c_{\alpha}|^2)}{d(\sum_\alpha|c_{\alpha}|^2)}).
\end{eqnarray} where we used
$\frac{\partial(\sum_\alpha|c_{\alpha}|^2)}{\partial(|c_{\alpha_0}|^2)}=1$.
On the right hand side we have also
\begin{eqnarray}
\frac{\partial \sum_\alpha
g(|c_{\alpha}|^2)}{\partial(|c_{\alpha_0}|^2)}=\frac{dg(|c_{\alpha_0}|^2)}{d(|c_{\alpha_0}|^2)}.
\end{eqnarray} and therefore we have
\begin{eqnarray}
\frac{dg(\sum_\alpha
|c_{\alpha}|^2)}{d(\sum_\alpha|c_{\alpha}|^2)})=\frac{dg(|c_{\alpha_0}|^2)}{d(|c_{\alpha_0}|^2)}.
\end{eqnarray}
Applying a second time the partial derivative on  both sides  but
now for the variable $\alpha_1\neq\alpha_0$ we get
 \begin{eqnarray}
\frac{\partial^2 g(\sum_\alpha
|c_{\alpha}|^2)}{\partial(|c_{\alpha_1}|^2)\partial(|c_{\alpha_0}|^2)}=
\frac{d^2g(\sum_\alpha
|c_{\alpha}|^2)}{(d(\sum_\alpha|c_{\alpha}|^2))^2}=\frac{\partial
\frac{dg(|c_{\alpha_0}|^2)}{d(|c_{\alpha_0}|^2)}}{\partial(|c_{\alpha_1}|^2)}=0.
\end{eqnarray} This implies    the differential equation
\begin{eqnarray}
\frac{d^2g(x)}{dx^2}=0.
\end{eqnarray}
 which has the trivial general solution
\begin{eqnarray}
g(x)=Ax+B
\end{eqnarray} with $A$ and $B$ constants. However for the
probability  we must have $g(0)=0$ and therefore $B=0$. This leads
to the normalized probability law
\begin{eqnarray}
p(\alpha)=\frac{g(|c_{\alpha}|^2)}{\sum_\alpha
g(|c_{\alpha}|^2)}=\frac{|c_{\alpha}|^2}{\sum_\alpha
|c_{\alpha}|^2},
\end{eqnarray} which constitutes Born's rule.\\ 
\indent Can we go beyond that approach? Years ago I found a small book by Jean-Louis Destouches~\cite{book} which motivated a different axiomatic. This suggested me an other possible `proof'  which goes like that:
 We start from the postulate for the probability density: 
\begin{eqnarray}
\rho_\psi(\textbf{X},t)=f(|\psi(\textbf{X},t)|)=g(\ln{(|\psi(\textbf{X},tx)|)})\end{eqnarray}   where $f$ or $g$ are function to be determined. For this consider the independence postulate for two quantum factorizable systems
\begin{eqnarray}
g(\ln{(|\psi_1(\textbf{X}_1,t)\psi_2(\textbf{X}_2,t)|)})\nonumber\\=g(\ln{(|\psi_1(\textbf{X}_1,t)|)}+\ln{(|\psi_2(\textbf{X}_2,t)|)})\nonumber\\=g(\ln{(|\psi_1(\textbf{X}_1,t)|)})g(\ln{(|\psi_2(\textbf{X}_2,t)|)})\end{eqnarray}
that is
\begin{eqnarray}
g(x+y)=g(x)g(y).\end{eqnarray}
 Writing $F(x)=\ln{(g(x))}$ we get
\begin{eqnarray}
F(x+y)=F(x)+F(y).\end{eqnarray} This can be solved like for
Gleason's theorem  we first take the $x$ derivative and then the $y$
derivative and we get finally $d^2 F(x)/dx^2=0$ i.e.
\begin{eqnarray}
F(x)=Ax+B\end{eqnarray} with $A$ and $B$ constants.
 We therefore have
 \begin{eqnarray}
g(x)=e^Be^{Ax}\end{eqnarray} which implies the probability rule
\begin{eqnarray}
\rho_\psi(\textbf{X},t)=g(\ln{(|\psi(\textbf{X},t)|)})=e^Be^{A\ln{(|\psi(\textbf{X},t)|)}}=e^B|\psi(x\textbf{X},t)|^A.\nonumber\\
\label{drezet}
\end{eqnarray}
To get Born's rule we now work in the de Broglie-Bohm mechanics
framework and postulate that we have the conservation rule
\begin{eqnarray}
\partial_t|\psi(\textbf{X},t)|^2=-\sum_{i=1}^{i=N}\boldsymbol{\nabla}_i\cdot[\textbf{v}_i(\textbf{X},t)|\psi(\textbf{X},t)|^2]\label{current}
\end{eqnarray} where
\begin{eqnarray}
\textbf{v}_i(\textbf{X},t)=\frac{\hbar}{m_i}Im[\frac{\boldsymbol{\nabla}\psi(\textbf{X},t)}{\psi(\textbf{X},t)}]
\end{eqnarray}is the de Broglie-Bohm velocity ($i=1,...,N$. The de Broglie-Bohm mechanics implies that we have also
the probability conservation Eq~\ref{madelung2bis} and therefore  the quantum Liouville theorem Eq.~\ref{liouvilian} $\frac{d}{dt}f(\mathbf{X}(t),t)=0$, which is a total particle derivative in the sense of Lagrange.\\ 
\indent Combining with Eq.~\ref{drezet} we get
\begin{eqnarray}
f(\textbf{X},t)=e^B|\psi(\textbf{X},t)|^{A-2},
\end{eqnarray}
and therefore  we deduce
\begin{eqnarray}
\frac{d|\psi(\textbf{X},t)|^{A-2}}{dt}=(A-2)|\psi(\textbf{X},t)|^{A-3}\frac{d|\psi(\textbf{X},t)|}{dt}=0\label{proba3}.
\end{eqnarray}
However, from Eq.~\ref{current}
\begin{eqnarray}
\frac{d|\psi(\textbf{X},t)|^2}{dt}=-|\psi(\textbf{X},t)|^2\sum_{i=1}^{i=N}\boldsymbol{\nabla}_i\cdot
\textbf{v}_i(\textbf{X},t)\neq 0\end{eqnarray} unless $\sum_{i=1}^{i=N}\boldsymbol{\nabla}_i\cdot
\textbf{v}_i(\textbf{X},t)=0$.
Consequently, we must in general have $A=2$ in Eq.~\ref{proba3} and
we deduce
\begin{eqnarray}
\rho(\textbf{X},t)=\frac{|\psi(\textbf{X},t)|^2}{\int d^{3N}\textbf{X}
|\psi(\textbf{X},t)|^2},\label{drezet2}
\end{eqnarray} which is Born's rule. The `derivation 'we propose here (see also Ref.~\cite{note}) is very similar
in spirit from the derivation  of the canonical Boltzmann-Gibbs
distribution  \begin{eqnarray}
\rho_{\textrm{Bolt.}}(q,p)=\frac{e^{\frac{-H(q,p)}{k_BT}}}{Z}\label{stat}\end{eqnarray}
for independent systems. Indeed Eq.~\ref{stat} can be justified
(e.g. in Landau and Lifshitz~\cite{Landau}) from the hypothesis that
$\rho_{\textrm{Bolt.}}(q,p)=f(H(q,p))$ must only depend on the
Hamiltonian function (one of the 7 additive constants of
motion for non interacting systems) and from the independence
postulate \begin{eqnarray}
f(H_1(q_1,p_1))f(H_2(q_2,p_2))\nonumber\\=f(H_1(q_1,p_1)+H_2(q_2,p_2)).\end{eqnarray}
The symmetry rule $f(x+y)=f(x)f(y)$, sometimes called  the Cauchy
equation when written in the form $F(x+y)=F(x)+F(y)$, was also used
long ago by Maxwell in his elementary derivation of the distribution
baring his name  (i.e. in 1860) and it also plays a role in the application of the
famous Boltzmann's $H$-theorem near equilibrium, i.e., when one is
applying the detailed balance principle $f(x')f(y')=f(x)f(y)$.  In
the de Broglie-Bohm mechanics  the $\Psi(x,t)$ function plays
somehow the role of the Hamilton-Jacobi action $S(x,t)$  in
classical mechanics: it guides the particle and also constrains the
statistical properties of an ensemble of identical systems. The
function $F(x,t)=\ln{(|\Psi(x,t)|^2)}$ appears therefore as a
dynamical constraint like the Hamiltonian can act for the
Boltzmann-Gibbs distribution. The present derivation therefore only
emphasizes that very fundamental aspect of the de Broglie-Bohm
quantum mechanics. Actually, it is clear from this analogy with thermodynamics, using the statistical independence postulate,  that $\ln{f(\textbf{X},t)}$  in the pilot wave interpretation should be a linear function of the additive constants of motions, i.e., $\ln{f(\textbf{X},t)}=\alpha+\beta H(\textbf{X},t)+...$ where $\alpha$ and $\beta$ are constant and $H$ is the energy.  However, in the quantum world defined by the guidance condition of de Broglie  $H(t)$ is not in general an integral of motion due to the presence of the quantum potential $Q$.  Therefore, the most general quantum equilibrium  should be  $\ln{f(\textbf{X},t)}=\alpha$ corresponding to the micro-canonical ensemble. A canonical ensemble should be only possible for the Hamiltonian eigenstates. This reasoning is interesting since it emphasizes the strong analogy between classical thermo-statistics in the phase-space (with conserved measure $d\Gamma_t=dqdp$)  and the quantum Bohmian-statistics (with conserved measure $d\Gamma_t=|\Psi(\textbf{X},t)|^2d^{3N}\textbf{x}$).             
\section{Probability and Born's rule in the pilot wave formalism: The Fokker-Planck approach }
All the previous attempts of deriving Born's rule were motivated by some `natural' symmetry considerations on probability.  These strategies have a old respectable origin and pioneers like  Boltzmann   Bernoulli,  Maxwell, or Laplace (to give some famous names) used very often such deductions.   However,  Boltzmann was probably the first motivated by the work of Maxwell to define kinetic equations for justifying the famous `H' theorem for the increase of entropy in gas.   Here, we would like to try a similar way of deriving Born's law in the pilot wave approach.  The method is motivated by the initial work proposed by Bohm and Vigier in the 1950's~\cite{BohmVigier}. However, they used a stochastic approach involving a  `subquantum' dynamic that we do not consider here. In their approach (which was also advocated by de Broglie in his `hidden thermodynamics'~\cite{Jalons}) the random fluctuation of the subquantum fluid forces the system to reach the equilibrium even when the guidance velocity vanishes (for example in fundamental atomic  `S' state). For the present purpose, based on the pilot-wave framework, we will instead use the Fokker-Planck equation and the Einstein-von Smoluchowski deduction of Brownian motion (we will use only one spatial dimension but the derivation is general). \\
\indent  We start from the equation 
\begin{eqnarray}
\rho(x,t+\tau)=\int P(x,t+\tau|x',t)\rho(x',t)dx',\label{oned}
\end{eqnarray} which defines the relation between a probability density $\rho(x,t+\tau)$ at time $t$ and point $x$ in phase space and the probability  at earlier time $t$ at various points $x'$. The term $P(x,t+\tau|x',t)$ can thus be seen as a transition probability between two points (with $\int P(x,t+\tau|x',t)dx=1$). 
and write
\begin{eqnarray}
P(x,t+\tau|x',t)=\int \delta(y-x)P(y,t+\tau|x',t).
\end{eqnarray}
Using the expansion
\begin{eqnarray}
\delta(y-x)=\delta(y-x'+x'-x)=\sum_{n=0}^{\infty}\frac{(y-x')^n}{n!}\frac{\partial^n}{\partial
x'^n}\delta(x'-x)
\end{eqnarray}
we obtain
\begin{eqnarray}
P(x,t+\tau|x',t)=\sum_{n=0}^{\infty}\frac{(-1)^n}{n!}
\cdot\frac{\partial^n}{\partial
x^n}[\int(y-x')^nP(y,t+\tau|x',t)\delta(x'-x)dy]\nonumber\\
=\sum_{n=0}^{\infty}\frac{(-1)^n}{n!}
\cdot\frac{\partial^n}{\partial
x^n}[\int(y-x)^nP(y,t+\tau|x,t)\delta(x'-x)dy]\nonumber\\
\end{eqnarray} which leads to
\begin{eqnarray}
\rho(x,t+\tau)-\rho(x,t)=\sum_{n=1}^{\infty}\frac{(-1)^n}{n!}
\cdot\frac{\partial^n}{\partial
x^n}[\int(y-x)^nP(y,t+\tau|x,t)\rho(x,t)dy].\nonumber\\
\end{eqnarray}From this we deduce in the limit  $\tau\rightarrow 0$  the well-know
Fokker-Planck formula corresponding to the so called Kramers-Moyal
expansion:
\begin{eqnarray}
\partial_t\rho(x,t)=\lim_{\tau\rightarrow 0}[\frac{\rho(x,t+\tau)-\rho(x,t)}{\tau}]\nonumber\\
=\sum_{n=1}^{\infty}(-1)^n\frac{\partial^n}{\partial
x^n}[D_n(x,t)\rho(x,t)dy],\label{moyal}
\end{eqnarray}
where the `diffusion constants', which can actually depend on $x$
and $t$, are defined as
\begin{eqnarray}
D_n(x,t)=\lim_{\tau\rightarrow
0}[\frac{1}{\tau}\int\frac{(y-x)^n}{n!}P(y,t+\tau|x,t)dy].
\end{eqnarray}
The usual approximation consists in neglecting terms   for $n>2$ and
we write
\begin{eqnarray}
\partial_t\rho(x,t)=-\frac{\partial}{\partial
x}[\bar{v}(x,t)\rho(x,t)]+D_t\frac{\partial^2}{\partial
x^2}\rho(x,t),
\end{eqnarray} where  $\bar{v}(x,t)=\lim_{\tau\rightarrow
0}[\int\frac{(y-x)}{\tau}P(y,t+\tau|x,t)dy]$ defines the average
velocity  and $D_t=\lim_{\tau\rightarrow
0}[\frac{1}{\tau}\int\frac{(y-x)^2}{2}P(y,t+\tau|x,t)dy]$ defines
the diffusion constant of the medium. The deterministic limit
corresponds to
\begin{eqnarray}
P(y,t+\tau|x,t)=\delta(y-X(t+\tau|x,t))
\end{eqnarray} where $X(t+\tau|x,t)$ denotes the position of the particle at the time $t+\tau$ linked by a deterministic trajectory
law to the earlier position $x$ of the same particle at time $t$. If
we insert this in Eq.~\ref{moyal} we get  $D_1=\lim_{\tau\rightarrow
0}[\frac{(X(t+\tau|x,t)-x)}{\tau}]=v(t)$  and $D_n=0$ for $n>1$.
This is in agreement with probability conservation and Liouville's
theorem for the de Broglie-Bohm mechanics (see Eq.~\ref{madelung2bis}).\\
\subsection{ Diffusion equation for a system coupled to a bath: relaxation to equilibrium}
As a  next step we consider a system S coupled to a thermostat T
with configuration coordinates labeled respectively as $x_S$ and
$x_T$. We write
\begin{eqnarray}
\rho_{S+T}(x_S,x_T,t+\tau)=\int\int
P(x_S,x_T,t+\tau|x_S',x_T',t)\nonumber\\
\cdot\rho(x_S',x_T',t)dx_S'dx_T'.
\end{eqnarray}
We will suppose  that at time $t$ the whole S+T system can
factorized as
\begin{eqnarray}
\rho_{S+T}(x_S',x_T',t)\simeq\rho_S(x_S',t)\rho_T(x_T',t)\label{factorization}\end{eqnarray} where
the thermostat state is at quantum equilibrium:
\begin{eqnarray}
\rho_T(x_T',t)=f_T(x_T',t)|\psi_T(x_T',t)|^2\simeq |\psi(x_T',t)|^2.
\end{eqnarray}
These relations are reminiscent from the old  factorization hypothesis (i.e. the \textit{Stosszahlansatz}) used by Boltzmann in his kinetic theory. 
This axiomatic allows us to write the reduced probability density
\begin{eqnarray}
\rho_S(x_S,t+\tau)=\int\rho_{S+T}(x_S,x_T,t+\tau)dx_T\nonumber\\=
\int Q(x_S,t+\tau|x_S',t)\nonumber\\
\cdot\rho_S(x_S',t)dx_S',\label{reat}
\end{eqnarray}
where
\begin{eqnarray}
Q(x_S,t+\tau|x_S',t)=\int\int
P(x_S,x_T,t+\tau|x_S',x_T',t)\nonumber\\
\cdot\rho_T(x_T',t)dx_Tdx_T'.\end{eqnarray} We point out that the
normalization conditions  $\int\int
P(x_S,x_T,t+\tau|x_S',x_T',t)dx_Sdx_T=1$ and $\int
\rho_T(x_T',t)dx_T'=1$ implies the normalization 
\begin{eqnarray}\int
Q(x_S,t+\tau|x_S',t)dx_S=1.\label{normalit}
\end{eqnarray} Therefore, the analogy with the
previous Eq.~\ref{oned} is complete and we can as well derive a
Fokker-Planck equation for the reduced system:
\begin{eqnarray}
\partial_t\rho_S(x_S,t)\simeq-\frac{\partial}{\partial
x}[\bar{v}_S(x_S,t)\rho_S(x_S,t)]\nonumber\\+D_t\frac{\partial^2}{\partial
x_S^2}\rho(x_S,t),\label{nonequili}
\end{eqnarray} where the diffusion parameters are defined as before.
In particular \begin{eqnarray}
\bar{v}_S(x_S,t)=\lim_{\tau\rightarrow
0}[\int\frac{(y_S-x_S)}{\tau}Q(y_S,t+\tau|x_S,t)dy_S]
\nonumber\\
=\lim_{\tau\rightarrow
0}[\int\int\int\frac{(y_S-x_S)}{\tau}P(y_S,x_T,t+\tau|x_S,x_T',t)\nonumber\\
\cdot\rho_T(x_T',t)dy_Sdx_T'dx_T].
\end{eqnarray}
However, the dynamic is deterministic therefore we have:
\begin{eqnarray}
P(y_S,x_T,t+\tau|x_S,x_T',t)\nonumber\\=\delta(y_S-x_0(t+\tau|x_s,x_T',t))
\nonumber\\\cdot\delta(x_T-X_0(t+\tau|x_S,x_T',t)),
\end{eqnarray} where $x_0(t+\tau|x_S,x_T',t)$ and
$X_0(t+\tau|x_s,x_T',t)$ represent the trajectories for the
particles. We therefore get
\begin{eqnarray}
\bar{v}_S(x_S,t)=\lim_{\tau\rightarrow
0}[\int\frac{(x_0(t+\tau|x_S,x_T',t)-x_S)}{\tau}\rho_T(x_T',t)dx_T']\nonumber\\
\end{eqnarray}
which can equivalently be written: \begin{eqnarray}
\bar{v}_S(x_S,t)=\frac{\int
v_S(x_S,x_T',t)\rho_{S+T}(x_S,x_T',t)dx_T'}{\int
\rho_{S+T}(x_S,x_T',t)dx_T'},
\end{eqnarray} with $v_S(x_S,x_T',t)=\nabla_SS(x_S,x_T',t)/m_S$ is
the pilot wave velocity for the particles belonging to the S
subsystem.
 We now suppose that we can write\begin{eqnarray}
\rho_S(x_S,t)=f_S(x_S,t)|\psi_S(x_S,t)|^2.
\end{eqnarray}
In the particular case where quantum equilibrium occurs
$f_S(x_S,t)=1$ and we can straightforwardly obtain:
\begin{eqnarray}
\partial_t|\psi_S(x_S,t)|^2\simeq-\frac{\partial}{\partial
x}[\bar{v}_S(x_S,t)|\psi_S(x_S,t)|^2]\nonumber\\+D_t\frac{\partial^2}{\partial
x_S^2}|\psi_S(x_S,t)|^2.\label{equili}
\end{eqnarray}
Using Eqs.~\ref{nonequili} and \ref{equili} we therefore obtain:
\begin{eqnarray}
\partial_tf_S(x_S,t)+\bar{v}_S(x_S,t)\cdot\nabla_S f_S(x_S,t)\nonumber\\=D_t[\nabla_S^2f_S(x_S,t)+2\nabla_S f_S(x_S,t)\cdot\frac{\nabla_S |\psi_S(x_S,t)|^2}{|\psi_S(x_S,t)|^2}].\label{fu}
\end{eqnarray}
This equation has solutions which converge to equilibrium. To see
this first observe that if at time $t$ we have an extremum $\nabla_S
f_S(x_S,t)=0$ then we deduce
\begin{eqnarray}
\partial_tf_S(x_S,t)=D_t\nabla_S^2f_S(x_S,t).\label{extrem}
\end{eqnarray}
For a local maximum we have $\nabla_S^2f_S(x_S,t)<0$ and therefore
$\partial_tf_S(x_S,t)<0$. Oppositely, for a local minimum we have
$\nabla_S^2f_S(x_S,t)>0$ and therefore $\partial_tf_S(x_S,t)>0$.  We
conclude that  local maxima will decay and local minima will
increase until $f_S(x_S,t)$ will reach a constant value (a similar conclusion was obtained by Bohm and Vigier using a stochastic alternative to the deterministic pilot wave theory of de Broglie~\cite{BohmVigier,Hiley}). This result  could be also  obtained if we solve  directly the diffusion equation in analogy with Schrodinger equation (i.e., by  using a complex time $it$). We  would deduce that any inhomogeneity are damped in time using a Green function $\frac{1}{\sqrt{4\pi(t-t')}^{3N}}e^{-\frac{(x-x')^2}{4D(t-t')}}$ with $3N$ the dimension of the phase space.  
\subsection{Equilibrium relaxation and the H theorem}
A different derivation is obtained  if we start from the integral
condition
\begin{eqnarray}
\frac{d}{dt}\int \rho_S(x_S,t) dx_S=\int
\partial_t\rho_S(x_S,t) dx_S\nonumber\\=-\int\nabla_S(\bar{v}(x_S,t)\rho_S(x_S,t)) dx_S\nonumber\\+D_t\int\nabla_S^2\rho_S(x_S,t)
dx_S\nonumber\\=-\oint\bar{v}_S(x_S,t)\rho_S(x_S,t)) \cdot
d\sigma_S\nonumber\\+D_t\oint\nabla_S\rho_S(x_S,t)\cdot
d\sigma_S=0\label{normus}
\end{eqnarray} where we used Gauss's theorem in the last step to
transform volume integrations into surface integration on the
infinitely remote boundary at which $\rho_S(x_S,t)$ and its spatial
derivative vanish. In order to use such conditions we introduce
Valentini's $H(t)$ functional~\cite{Valentini} which, in analogy with Boltzmann's  $H
$ reads:
\begin{eqnarray}
H(t)=\int f_S(x_S,t)\ln{[f_S(x_S,t)]}|\psi_S(x_S,t)|^2dx_S.
\end{eqnarray}
The time derivative of $H$ is given explicitly by
\begin{eqnarray}
\frac{d}{dt}H=\int[
\ln{(f_S)}\partial_t(f_S|\psi_S|^2)+|\psi_S|^2\partial_tf_S]dx_S
\nonumber\\
=\int[-\ln{(f_S)}\nabla_S(f_S|\psi_S|^2\bar{v}_S)+D_t\ln{(f_S)}\nabla_S^2(f_S|\psi_S|^2)\nonumber\\
-\bar{v}|\psi_S|^2\partial_tf_S+2D_t\nabla_Sf_S\nabla_S|\psi_S|^2+D_t|\psi_S|^2\nabla_S^2f_S]dx_S\nonumber
\end{eqnarray} where we used  Eqs.~\ref{nonequili} and \ref{equili}.
After an integration by part and some rearrangements we can rewrite
$dH/dt$ as
\begin{eqnarray}
\frac{d}{dt}H
=\int[-\nabla_S(\ln{(f_S)}f_S|\psi_S|^2\bar{v}_S)\nonumber\\+D_t\nabla_S^2(\ln{(f_S)}f_S|\psi_S|^2)
 -D_t|\psi_S|^2\frac{(\nabla_S f_S)^2}{f_S}]dx_S
\end{eqnarray} where the two first terms in the right hand side
vanish as in Eq.~\ref{normus} after using Gauss's theorem and
boundary conditions at infinity. We therefore deduce
\begin{eqnarray}
\frac{d}{dt}H =-D_t\int|\psi_S|^2 \frac{(\nabla_S f_S)^2}{f_S}dx_S\leq0
\end{eqnarray} which is a form of $H$-theorem for our diffusive  de Broglie-Bohm model. We immediately conclude that in order to reach a minimum for
$H(t)$ canceling $dH/dt$  at time $t_{eq}$ we must necessarily have
$\nabla_S f_S=0$ everywhere. From Eq.~\ref{extrem} this will
necessarily lead to $\partial_tf_S=0$ at that time and in the future
of $t_{eq}$.\\ 
\indent The calculation presented here can be also expressed using a master equation similar to the one used by Boltzmann to derive the H-theorem.  For this we start from Eq.~\ref{reat} written as $f_S(x_S,t+\tau)=
\int \frac{Q(x_S,t+\tau|x_S',t)}{|\psi_S(x_S,t+\tau)|^2}f_S(x_S',t)|\psi_S(x_S',t)|^2dx_S'$  and define the partial time derivative \begin{eqnarray}\partial_t f_S(x_S,t)=\lim_{\tau\rightarrow
0}[\frac{f_S(x_S,t+\tau)-f_S(x_S,t)}{\tau}].\label{vitessesup}\end{eqnarray} We thus obtain (using the normalization Eq.~\ref{normalit}) the master equation:  
 \begin{eqnarray}
\partial_t f_S(x_S,t)=\int dx_S' |\psi_S(x_S',t)|^2[\frac{K(x_S,x_S',t)}{|\psi_S(x_S,t)|^2}f_S(x_S',t)\nonumber\\-\frac{K(x'_S,x_S,t)}{|\psi_S(x'_S,t)|^2}f_S(x_S,t)]\label{master}
\end{eqnarray} with $K(x_S,x_S',t)=\lim_{\tau\rightarrow
0}\frac{Q(x_S,t+\tau|x_S',t)}{\tau}\geq 0$ and $Q(x_S,t|x_S',t)=\delta(x_S-x_S')$. Now, we define the conditional $H$ function as 
\begin{eqnarray}H_r(t)=\int dx_S |\psi_S(x_S,t)|^2f_S(x_S,t)\ln{(\frac{f_S(x_S,t)}{g_S(x_S)})}\end{eqnarray} where $g_S(x_S)$ is a	 stationary solution of Eq.~\ref{master} (i.e. with $\partial_t g_S(x_S,t)=0$ for all time). With such a definition, and using Eq.~\ref{master} together with the Liouville conservation of the phase volume with time (i.~e. $\frac{d}{dt}\delta x_S |\psi_S(x_S,t)|^2$), it is easy to derive  the equality (see \cite{Prigogine2}):
  \begin{eqnarray}
\frac{d}{dt}H_r(t)=\int \int d\Gamma_S(x_S,t)d\Gamma_S(x_S',t)\frac{K(x_S,x_S',t)}{|\psi_S(x_S,t)|^2}g_S(x'_S)\nonumber\\ \cdot\frac{f_S(x'_S,t)}{g_S(x'_S)}\ln{(\frac{f_S(x_S,t)g_S(x'_S)}{f_S(x'_S,t)g_S(x_S)})}.\end{eqnarray} Finally, using the inequality $\ln{(\frac{f_S(x_S,t)g_S(x'_S)}{f_S(x'_S,t)g_S(x_S)})}\leq \frac{f_S(x_S,t)g_S(x'_S)}{f_S(x'_S,t)g_S(x_S)}-1$ and the definition of the stationary state we deduce again the H-theorem: \begin{eqnarray}\frac{d}{dt}H_r(t)\leq \int \int d\Gamma_S(x_S,t)d\Gamma_S(x_S',t)[\frac{K(x_S,x_S',t)}{|\psi_S(x_S,t)|^2}g_S(x_S')\nonumber\\-\frac{K(x_S',x_S,t)}{|\psi_S(x_S',t)|^2}g_S(x_S)]\frac{f_S(x_S,t)}{g_S(x_S)}=0,\end{eqnarray} i.e., an irreversible increase of relative entropy $\frac{d}{dt}S_r(t)=-\frac{d}{dt}H_r(t)\geq 0$ in agreement with the second law of thermodynamic (applied here to the pilot wave interpretation). Importantly, the equality occurs only if $\frac{f_S(x_S,t)g_S(x'_S)}{f_S(x'_S,t)g_S(x_S)}=1$. But since the master equation possesses by hypothesis the simple stationary solution $g_S(x_S)=1$ we have at equilibrium necessarily $\frac{f_S(x_S,t)}{f_S(x'_S,t)}=1$  meaning that the quantum equilibrium is unique.  Clearly, the H-theorem breaks time symmetry and that means that the Master equation should be considered with caution. Indeed from the definition we have $K(x_S,x_S',t)\geq 0$ only if $\tau>0$.  The previous H-theorem  is thus only valid near   equilibrium   for $\tau>0$. In order to reestablish  the time symmetry we consider the alternative time derivative $\partial_t f_S(x_S,t):=\lim_{\tau\rightarrow
0}[\frac{f_S(x_S,t-\tau)-f_S(x_S,t)}{-\tau}]$ which in the continuous limit is intuitively the same as the previous definition Eq.~\ref{vitessesup}. However, we now deduce the Kinetic equation:
     \begin{eqnarray}
\partial_t f_S(x_S,t)=\int dx_S' |\psi_S(x_S',t)|^2[\frac{K'(x_S,x_S',t)}{|\psi_S(x_S,t)|^2}f_S(x_S',t)\nonumber\\-\frac{K'(x'_S,x_S,t)}{|\psi_S(x'_S,t)|^2}f_S(x_S,t)]\label{master}
\end{eqnarray} with $K'(x_S,x_S',t)=\lim_{\tau\rightarrow
0}\frac{Q(x_S,t-\tau|x_S',t)}{-\tau}\leq 0$. Clearly, we can obtain again a kind of H theorem but since  $K'(x_S,x_S',t)\leq 0$ this actually means $\frac{d}{dt}H_r(t)\geq 0$ or equivalently stated $H_r(t-\tau)-H_r(t)\leq 0$ which implies a decrease of $H$ in the past! Of course  we assume that this kinetic or transport equation  makes sense only near the equilibrium point which is a kind of Markovian approximation where the typical time $\tau$ plays a crucial role for the dynamics. The results we obtained demonstrate, in agreement with a discussion given by Boltzmann in ref.\cite{Gas1}, that the equilibrium state is stable against perturbation ant that if the system eventually fluctuates near the distribution $f=1$ it will finally return to equilibrium both in the future and the past (see the discussion about the so called `H curves' by Boltzmann, Culverwell and Burbury \cite{Gas1}).    \\            
\indent The previous analysis should also be compared with the coarse-graining method proposed by Valentini~\cite{Valentini} for obtaining a version of the Gibbs-Tolman H-theorem. In this famous strategy a coarse-grained distribution  $\overline{f}$ is obtained after averaging on some degrees of freedom of the system. This allows Valentini to derive an inequality $\int_{\Gamma}
\overline{f}\ln{\overline{f}}d\Gamma\leq \int_{\Gamma}
f\ln{f}d\Gamma$ which, like in the Ehrenfests or Tolman reasoning, can be used to define a kind of time arrow and an irreversible tendency for reaching equilibrium. The well-known issue~\cite{Davies,Zeh} with this kind of proof is that it relies on the assumption that the inequality defines a hierarchy between an entropy at the initial time (where the fine grained distribution $f$ is used) and the entropy at a subsequent time (where the coarse-grained distribution $f$ is used). However, we have no reason to believe that this will be valid at all times so that the condition $\frac{d}{dt}[\int_{\Gamma}
\overline{f}\ln{\overline{f}}d\Gamma] \leq 0$  can not be rigorously inferred~\cite{Callender} as a dynamical constraint for all times. The strategy used in our work is however not completely orthogonal to this H-theorem proof since, like in the Valentini or Tolman work, we used some averaging on what we called  the thermostat $T$ (similar ideas are defended in a more usual framework by R. Balian \cite{Balian}). Obtaining a master equation or a diffusion equation somehow means indeed to add axioms or postulates for deriving an inequality valid at all times. This was clear with Boltzmann work where the molecular chaos condition played a key role for introducing irreversibility in a otherwise time symmetric dynamics. Here, the same occurs with our diffusive dynamics since the factorization condition Eq.~\ref{factorization} is very similar in spirit to the factorization hypothesis introduced by Boltzmann. Clearly the role of entanglement is central in our deduction since it is the averaging on the bath degrees of freedom, supposed to be already in equilibrium, which allows us to define a kind of master equation for the quantum fluid associated with the sub-system $S$.\\  
\indent The dynamical proof we considered in this paper is actually more similar in philosophy to the work proposed initially by  Bohm and Vigier \cite{BohmVigier}. However, They considered a stochastic approach with a suquantum dynamics. This idea has a very long tradition (see for example refs.~\cite{Furth}).  Here, however by emphasizing the role  of entropy  and of diffusion, we were able to analyze quantitatively this approach by using the Fokker-Planck formalism in a strict Bohmian context, i.e.,  without adding a subquantum dynamics.  What is also clear is that  we need to couple a system $S$ to a thermostat $T$ to obtain the proof. This is therefore a signature of the fundamental role of the environment in order to justify any kinetic equations.  This is well recognized in Boltzmann work and in modern quantum diffusion theory. By applying  some of these concepts to the pilot wave framework built by de Broglie we open new possibility for understanding how the quantum equilibrium can be reached and  how deviations can be observed.  It could be particularly important to connect this issue with the  strategy proposed, for example by Ilya Prigogine, for explaining the tendency to reach an equilibrium through deterministic chaos and mixing. The strategy was recently exploited in the context of  the pilot wave theory by H. Geiger, G. Obermair and Ch. Helm~\cite{Geiger2001} using the so called Bernoulli shift which is an example of chaotic deterministic map. The issue is however more complicated  than these authors thought at the beginning since coherence of the Schrodinger wave in general prohibits trajectories to cross. Therefore, in order to exploit the concept of chaotic map in the de Broglie Bohm  approach it was recognized that entanglement with an environment is a key ingredient to erase part of the coherence (i.e. due to the loss of information through entanglement allowing to transform a pure state into a mixture)~\cite{Philbin}.\\
\indent Consider for example a wave packet which in a  good approximation can be described as a plane-wave.  This wave packet is supposed to propagate along the $x$ direction uniformly.   Then, periodically this wave packet interacts adiabatically with a potential.  We are interested in the dynamics along the transverse direction $y$ and we suppose that the effect of the interaction is to contract the wave packet in a very small region transversely. Calling $\Phi_n(y)$ the wave function before the contraction we have after the interaction $\Phi'_{n}(y)=\sqrt{L}\Phi_{n}(Ly)$ where $L>1$. To fix the idea we suppose that  $\Phi_n(y)=1$ if $0<y<1$ and cancels otherwise. Therefore, we have $\Phi'_n(y)=\sqrt{L}$ if $0<y<1/L$ and $\Phi'_n(y)=0$ otherwise. Subsequently, we suppose that the  contracted wave packet interacts with an environment. The  coupling induces entanglement and we have at the end a wave function like $\Phi'_n(y)[\theta(y\cdot L-1/2))|\epsilon_{n,-}\rangle+\theta(-y\cdot L+1/2)|\epsilon_{n,+}\rangle]$ where $|\epsilon_{n,\pm}\rangle$ are orthogonal quantum states of the environment. Finally, after the entanglement the wave packet re-expands and we return to the initial size along the $y$ direction: $|\Phi_{n+1}\rangle=\Phi_n(y)[|\epsilon_{n,-}\rangle+|\epsilon_{n,+}\rangle]\sqrt{2}$. The square root of 2 for the normalization comes from the diffraction regime considered during this step (i.e. a diffraction regime where the two wave packets entangled with $|\epsilon_{n,\pm}\rangle$ evolves independently to occupy at the end the full available space $y\in [0,1]$). Thanks to entanglement the trajectories can now cross and we can easily find examples where the positions of the particle at the step $n$ and $n+1$ are related by the Bernoulli map: $y_{n+1}=2 y_n \textrm{(mod 1)}$.  This map is chaotic which means that  a very small difference in the initial positions will grow exponentially with $n$ (the Lyapounoff exponent is $\ln{2}$). From the point of view of the density of probability what we will get is a Perron-Frobenius iterative relation~\cite{Prigogine,Birula} such as:
\begin{eqnarray}
\rho_{n+1}(y)=\frac{1}{2}(\rho_{n}(\frac{y}{2})+\rho_{n}(\frac{y+1}{2})).
\end{eqnarray} This is very interesting since one can directly see that  $\rho_n(y)\rightarrow 1$ for $n\rightarrow+\infty$ whatever the initial conditions for $\rho_0(y)$ (at least if the distribution is not singular). This is because the density can be expanded using Bernoulli polynomials $B_m(y)$ as
\begin{eqnarray}
\rho_{n}(y)=\sum_{m=0}^{m=+\infty}C_m e^{-nm\textrm{ln}(2)}B_m(y).
\end{eqnarray}
and for $n\rightarrow +\infty$ only the constant term $B_0(1)$ with coefficient $C_0=1$ survives.\\
  This implies that any inhomogeneities in density of probability will get damped with time and will reach a quantum equilibrium, i.e., in general very quickly after only few iterations.
The process is very similar in spirit to the previous dissipative model we proposed and it could be interesting to analyze further the  role of entanglement in the relaxation to equilibrium.
 Observe also that in the asymptotic regime where $\rho_{n}(y)\simeq 1+C_1 e^{-\textrm{ln}(2)}B_1(y)$ we have 
\begin{eqnarray}
\rho_{n+1}(y)-\rho_{n}(y)=-(\rho_{n}(y)-1)(1-e^{-\textrm{ln}(2)}).
\end{eqnarray} which can be written in the continuous limit
\begin{eqnarray}
\frac{d}{dt}\rho_{t}(y):=-\frac{(\rho_{t}(y)-1)}{\tau}.
\end{eqnarray} where we introduced the time $\tau=\tau_0(1-e^{-\textrm{ln}(2)})^{-1}$ and $\tau_0$ is a typical cycle time for the contraction expansion process. This formula is reminiscent of the collision term sometimes used with Boltzmann's equation so that a relation is here sketched between the chaotic (i.e. a la Prigogine) and diffusive (i.e. a la Boltzmann) approach. This should motivate further work on this topic.\\   
 Finally, as it was pointed out years ago by Bohm~\cite{Hiley}, deterministic chaos near phase singularities should play an important role in this process~\cite{Chaos} since it will act as a source for relaxation with more realistic models than the paradigmatic Bernoulli map used before. Bohm in 1953~\cite{Bohm2} already showed examples where correlations with the environment during a scattering process lead to relaxation to quantum equilibrium.  Clearly, more studies should  be done to understand this kinetic relaxation regime precisely (and its relation with mixing,  K-system, or Bernoulli system). This will be a central issue in order to see how robust is quantum equilibrium in the pilot wave approach.   
                            
\section{Conclusion}
 In this paper we presented several derivations or proofs of Born's rule in the pilot wave framework. All these approaches (and some not discussed in this article~\cite{Callender}) based on symmetry  considerations or dynamical modeling  lead to the conclusion that the quantum equilibrium is a very natural consequence of the pilot wave theory (if we add some intuitive axioms concerning initial conditions in the Universe). We think that this issue  should motivate other researches, like  the ones A. Valentini attempted in the recent years~\cite{Valentini,Valentini2} to find a regime of quantum non-equilibrium where Born's rule could fail and where ultimately the pilot wave theory will differ from standard quantum mechanics. \\
Finally, the main issue for future works concern the double solution program of de Broglie. We believe that it is possible to construct a version of de Broglie theory exempts of any contradiction such as it reproduces completely  the predictions of the usual quantum formalism as well as the trajectories of the pilot wave  dynamics of de Broglie and Bohm (without adding subquantum forces).  This will be developed in a future article.    
\section{Acknowledgments} 
We would like to thank the organizers of the FQXI Symposium in Marseille 2016 for bringing us the opportunity to discuss our work.   In particular we would like to thank T. Durt for motivating the redaction of the present article. I also kindly thank W. Struyve for bringing to my knowledge the paper given in Ref.\cite{note} after the present manuscript was written.   


\end{document}